\documentclass[letterpaper]{article} 
\usepackage{aaai23}  
\usepackage{amsmath,amsfonts,amssymb}
\usepackage{times}  
\usepackage{helvet}  
\usepackage{courier}  
\usepackage[hyphens]{url}  
\usepackage{graphicx} 
\def\UrlFont{\rm}  
\usepackage{natbib}  
\usepackage{caption} 
\usepackage{algorithm}
\usepackage{algorithmic}
\usepackage{url}
\usepackage{enumitem}
\usepackage{colortbl}
\usepackage{multirow}
\usepackage{makecell}
\usepackage{scalerel}
\usepackage{bigstrut}
\usepackage[hidelinks]{hyperref}
\usepackage{pifont}
\newcommand{\xmark}{\ding{55}}%

\newlength\savedwidth

\newlength\savewidth
\newcommand\shline{\noalign{\global\savewidth\arrayrulewidth
                            \global\arrayrulewidth 1.5pt}%
                   \hline
                   \noalign{\global\arrayrulewidth\savewidth}}
\newcolumntype{"}{@{\hskip\tabcolsep\vrule width 1pt\hskip\tabcolsep}}
\definecolor{mygray}{gray}{0.5}
\definecolor{background_gray}{gray}{0.9}

\definecolor{commentcolor}{RGB}{110,154,155}   

\usepackage{newfloat}
\usepackage{listings}
\DeclareCaptionStyle{ruled}{labelfont=normalfont,labelsep=colon,strut=off} 
\floatstyle{ruled}
\newfloat{listing}{tb}{lst}{}
\floatname{listing}{Listing}
\title{AirFormer: Predicting Nationwide Air Quality in China with Transformers}
\author{
    Yuxuan Liang\textsuperscript{\rm 1},
    Yutong Xia\textsuperscript{\rm 1},
    Songyu Ke\textsuperscript{\rm 4,2},
    Yiwei Wang\textsuperscript{\rm 1},
    Qingsong Wen\textsuperscript{\rm 3} \\
    Junbo Zhang\textsuperscript{\rm 2},
    Yu Zheng\textsuperscript{\rm 2},
    Roger Zimmermann\textsuperscript{\rm 1}
}
\affiliations{
    \textsuperscript{\rm 1}National University of Singapore \textsuperscript{\rm 2}JD Intelligent Cities Research \& JD iCity, JD Technology \\
    \textsuperscript{\rm 3}DAMO Academy, Alibaba Group \textsuperscript{\rm 4}Shanghai Jiao Tong University\\
    \{yuxliang, songyu-ke,msjunbozhang, msyuzheng\}@outlook.com; \\ \{xiayutong618,qingsongedu\}@gmail.com;  \{y-wang,rogerz\}@comp.nus.edu.sg 
    
}

\usepackage{bibentry}

\begin{document}

\maketitle

\begin{abstract}
Air pollution is a crucial issue affecting human health and livelihoods, as well as one of the barriers to economic and social growth. Forecasting air quality has become an increasingly important endeavor with significant social impacts, especially in emerging countries like China. In this paper, we present a novel Transformer architecture termed AirFormer to collectively predict nationwide air quality in China, with an unprecedented fine spatial granularity covering thousands of locations. AirFormer decouples the learning process into two stages -- 1) a bottom-up deterministic stage that contains two new types of self-attention mechanisms to efficiently learn spatio-temporal representations; 2) a top-down stochastic stage with latent variables to capture the intrinsic uncertainty of air quality data. We evaluate AirFormer with 4-year data from 1,085 stations in the Chinese Mainland. Compared to the state-of-the-art model, AirFormer reduces prediction errors by 5\%$\sim$8\% on 72-hour future predictions. Our source code is available at \url{https://github.com/yoshall/airformer}.
\end{abstract}

\section{Introduction}
Air pollution refers to the release of air pollutants, such as gases, dust, fumes, and odors, into the atmosphere that are detrimental to human health and the environment. According to the World Health Organization (WHO), air pollution is one of the leading causes of death in the world today, accounting for seven million fatalities per year \cite{vallero2014fundamentals}. Over 90\% of people breathe air that contains more contaminants than the WHO's recommended levels, with those in emerging countries suffering the most, especially in China with 1.4 billion people \cite{wang2012air}.

To inform citizens about real-time air quality, many Chinese cities have built a number of air quality monitoring stations. These stations report time series readings every hour, including the concentration of particulate matter (PM$_{2.5}$ and PM$_{10}$), NO$_{2}$, etc. For instance, a 50 µg/m$^3$ concentration of PM$_{2.5}$ refers to good air quality, indicating ``a terrific day to be active outside'' according to the definition of the U.S. Environmental Protection Agency \cite{lin2018exploiting}. Beyond real-time monitoring, forecasting air quality also becomes an increasingly important endeavor with the economic, ecologic, and human toll that air pollution takes, which significantly aids human health protection (e.g., informing people whether to travel outdoor) and government policy-making.

\begin{figure}[!t]
  \centering
  \includegraphics[width=0.47\textwidth]{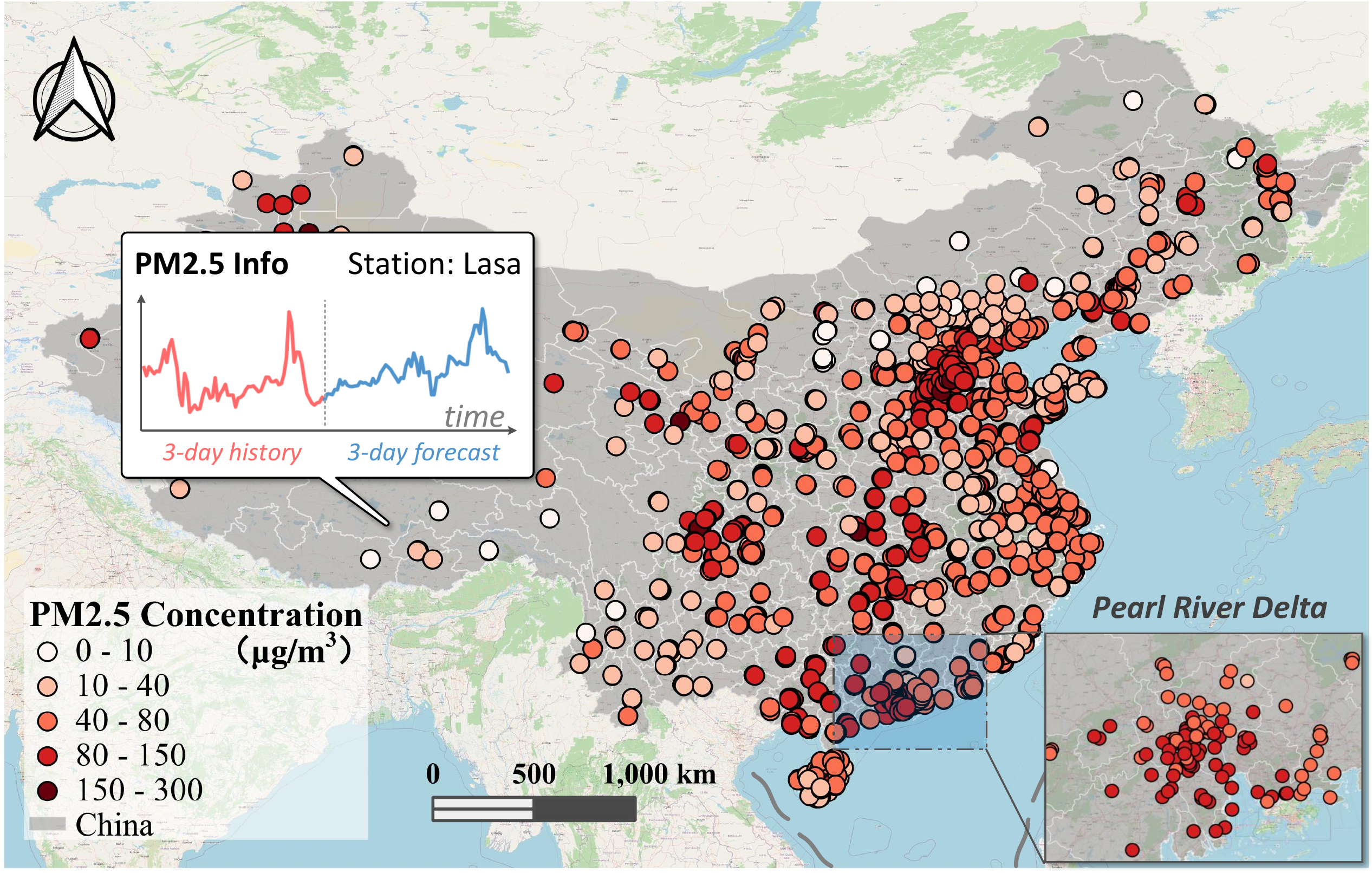}
  \vspace{-1em}
  \caption{Distribution of 1,085 air quality stations in China. } 
  \label{fig:intro}
  \vspace{-1.5em}
\end{figure}

During the past decades, there has been a long-established line of research conducted for air quality prediction, ranging from classical dispersion models \cite{vardoulakis2003modelling} to data-driven models \cite{zheng2015forecasting,yi2018deep}. Considering the computational expense, they mostly focused on predicting the air quality at a city scale with dozens of monitoring stations, e.g., there are 35 stations in Beijing. With recent advances in deep learning, researchers started to explore nationwide air quality prediction, i.e., \emph{collectively} predict on nearly two hundred stations in China using Spatio-Temporal Graph Neural Networks (STGNNs) \cite{wang2020pm25,chen2021group}. STGNNs couple Graph Neural Networks (GNNs) with deep sequential models, e.g., Recurrent Neural Networks (RNNs), in which GNNs are used to capture the spatial correlations among stations (i.e., dispersion), and RNNs are utilized to learn the temporal dependencies. 

Further, attention-based models, in particular Transformers, have become a strong alternative to capture spatial correlations in air quality data \cite{wang2021modeling,wang2022air}. They have two major merits over STGNNs. First, they jointly capture the short-term and long-term interactions among different places at each layer, while STGNNs merely convolve the local surroundings. Second, the correlation of air quality between different locations is highly dynamic, changing over time \cite{liang2018geoman,cheng2018neural}. Using attention-based models can naturally address this issue.

In this study, we broaden our scope to collectively predict air quality in Chinese mainland with an unprecedented fine spatial granularity using transformers, covering thousands of stations. As shown in Figure \ref{fig:intro}, our prediction targets encompass all provinces of the Chinese mainland, with dense distribution in developed regions, e.g., Pearl River Delta. Such a fine coverage not only provides more useful information to the public with high social impacts, but also comprises more data samples that benefit model training \cite{zhao2015multi}. 

Despite these benefits, forecasting air quality with a fine spatial granularity poses a significant modeling difficulty for transformers -- \emph{efficiency}. Multi-head Self-Attention (MSA), which is the key operation of transformers for spatial modeling, takes quadratic computational complexity w.r.t the number of stations $N$. Such expense may become unaffordable with the growth of $N$, especially for our fine-grained data. 

Meanwhile, the future air quality readings are intrinsically \emph{uncertain} due to two factors -- inaccurate or missing observations, and some unpredictable factors, e.g., vehicle exhaust, policy, and industrial emission. While earlier attempts have shown promising performance on air quality prediction via deterministic approaches, most of them still fall short of capturing such uncertainty within large-scale air quality data.

\begin{figure}[!t]
  \centering
  \includegraphics[width=0.47\textwidth]{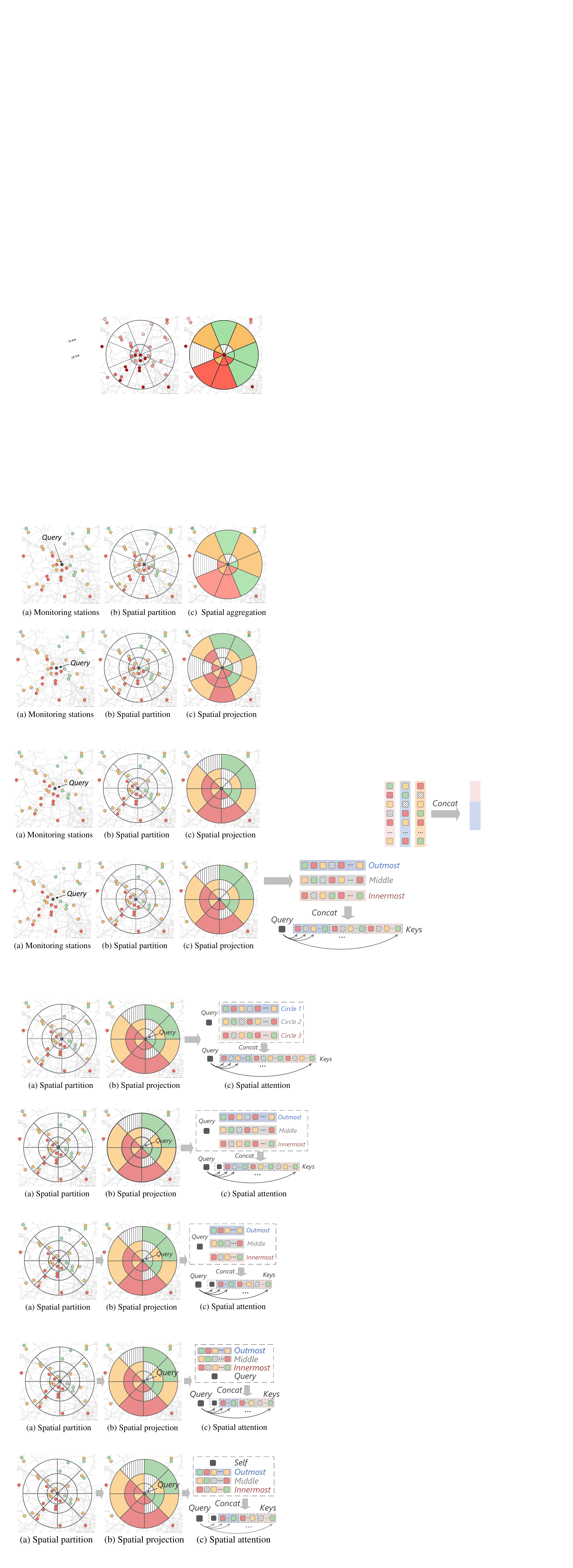}
  \vspace{-1.5em}
  \caption{Sketch of DS-MSA. For a query location (the black point), we first divide its surroundings into regions bounded by three circles and four lines. Then we project other stations onto the regions to obtain regional representations. Finally, we use the station feature as the query and the regional representations as the keys and values for attention computation.} 
  \label{fig:dartboard}
  \vspace{-1em}
\end{figure}

To tackle these challenges, we present a novel transformer architecture for nationwide air quality prediction in China, entitled \textbf{AirFormer}. Our method is motivated by the domain knowledge of air pollution, which enables us to build models with more interpretations. AirFormer decouples the solution to the two issues into dual stages -- a \emph{deterministic} stage and a \emph{stochastic} stage. In the deterministic stage, we propose two new types of MSA to \emph{efficiently} capture the spatial and temporal dependencies, respectively. In the stochastic stage, we explore the inclusion of \emph{latent random variables} into the transformer. These latent variables are sampled from probability distributions that are learned from the deterministic hidden states, thus capturing the uncertainty of the input data. In summary, our contributions lie in four aspects:
\begin{itemize}[leftmargin=*]
    \vspace{-0.2em}
    \item Considering that the spatial correlations among nearby locations are often stronger than those far away, we devise the \emph{Dartboard Spatial MSA} (DS-MSA) to efficiently capture spatial relations. As its name suggests, each location attends to its close surroundings at a fine granularity and faraway stations at a coarse granularity (see Fig. \ref{fig:dartboard}). Compared to standard MSA with quadratic cost, DS-MSA only takes \emph{linear complexity} w.r.t the number of stations.
    
    \vspace{-0.1em}
    \item We devise \emph{Causal Temporal MSA} (CT-MSA) for learning temporal dependencies. It ensures that the output of a step derives only from previous steps, i.e., causality. Locality is also introduced to improve efficiency, where the receptive field in each layer is gradually increased like convolutions.
    
    \vspace{-0.1em}
    \item Leveraging recent advances in variational models, we enhance transformers with \emph{latent variables} to capture the uncertainty of air quality data. To preserve the parallelism of transformers, the latent random variables are arranged hierarchically with implicit temporal dependencies.
    
    \vspace{-0.1em}
    \item To the best of our knowledge, this is the first work for \emph{collectively} forecasting air quality among thousands of locations. The empirical results show that AirFormer obtains 4.6\%-8.2\% lower prediction errors than existing models. 
\end{itemize}

\section{Preliminary}
\noindent \textbf{Problem Formulation}

\noindent The readings of $N$ air quality monitoring stations at a given time $t$ can be denoted as $\mathbf{X}_t \in \mathbb{R}^{N \times D}$, where $D$ is the number of measurements, including \emph{air pollutants} (e.g., PM$_{2.5}$, NO$_2$) and \emph{external factors} (e.g., weather, wind speed). Each entry $x_{ij}$ indicates the value of the $j$-th measurement of the $i$-th station. Given the historical readings of all stations from the past $T$ time steps, we aim to learn a function $\mathcal{F}(\cdot)$ that predicts $D^\prime$ kinds of measurements over the next $\tau$ steps: 
\vspace{-0.4em}
\begin{equation}
    \mathbf{X}_{1:T} \stackrel{\mathcal{F}(\cdot)}{\rightarrow} \mathbf{Y}_{1:\tau},
\end{equation}
where $\mathbf{X}_{1:T} \in \mathbb{R}^{T \times N \times D}$ is the history data and $\mathbf{Y}_{1:\tau} \in \mathbb{R}^{\tau \times N \times D^\prime}$ is the future readings.
\vspace{0.2em}

\noindent \textbf{Multi-head Self-Attention (MSA)}

\noindent MSA is the key operation of transformers to learn an alignment where each token in the sequence learns to gather messages from other tokens \cite{vaswani2017attention}. Let $\mathbf{X} \in \mathbb{R}^{S \times C}$ be the input sequence with length $N$ and feature dimension $C$. The operation of a single head is defined as:
\begin{equation}\label{eq:msa}
    \mathbf{X}_{h}=\operatorname{Softmax}\left( \alpha \mathbf{Q}_{h} \mathbf{K}_{h}^{\top} \right) \mathbf{V}_{h},
\end{equation}
where $\mathbf{X}_{h} \in \mathbb{R}^{S \times C/N_h}$ is the output features; $\mathbf{Q}_{h} = \mathbf{X} \mathbf{W}_{q}$, $\mathbf{K}_{h} = \mathbf{X} \mathbf{W}_{k}$ and $\mathbf{V}_{h} = \mathbf{X} \mathbf{W}_{v}$ are queries, keys, and values, respectively; $\mathbf{W}_{q}, \mathbf{W}_{k}, \mathbf{W}_{v} \in \mathbb{R}^{C \times C/N_h}$ are the learnable parameters for linear projection, and $N_h$ is the number of heads; $\alpha$ is a scaling factor. The computational complexity of Eq. (\ref{eq:msa}) is quadratic w.r.t the sequence length $S$.
\vspace{0.2em}

\noindent \textbf{Variational Models with Latent Variables}

\noindent Variational autoencoders (VAE) \cite{kingma2013auto} have long been verified to be an effective modeling paradigm for recovering complex multimodal distributions over the latent space. VAE addresses the data distribution $p(\mathbf{x})$ using an unobserved latent variable $\mathbf{z}$ and is parameterized by $\theta$ as:
\begin{equation}
p_\theta(\mathbf{x})=\scaleobj{.65}{\int} p_\theta(\mathbf{x} | \mathbf{z}) p_\theta(\mathbf{z}) d \mathbf{z}.
\end{equation}
As the integral is usually intractable, VAE introduces an approximate posterior $q_{\phi}(\mathbf{z} | \mathbf{x})$ and implicitly optimizes the evidence lower bound (ELBO) of the marginal log-likelihood:
\begin{equation}
\nonumber
    \log p_{\theta}(\mathbf{x}) \geq-K L\left(q_{\phi}(\mathbf{z} | \mathbf{x}) \| p_{\theta}(\mathbf{z})\right)+\mathbb{E}_{q_{\phi}(\mathbf{z} | \mathbf{x})}\left[\log p_{\theta}(\mathbf{x} | \mathbf{z})\right],
\end{equation}
where $KL$ denotes the KL divergence. The prior $p_\theta(\mathbf{z})$ and the posterior $q_{\phi}(\mathbf{z} | \mathbf{x})$ of the latent variables are usually taken to be Gaussian distribution with diagonal covariance, which inherently encodes the uncertainty of the input data.


\section{Methodology}
\begin{figure}[!b]
  \centering
  \vspace{-1.5em}
  \includegraphics[width=0.5\textwidth]{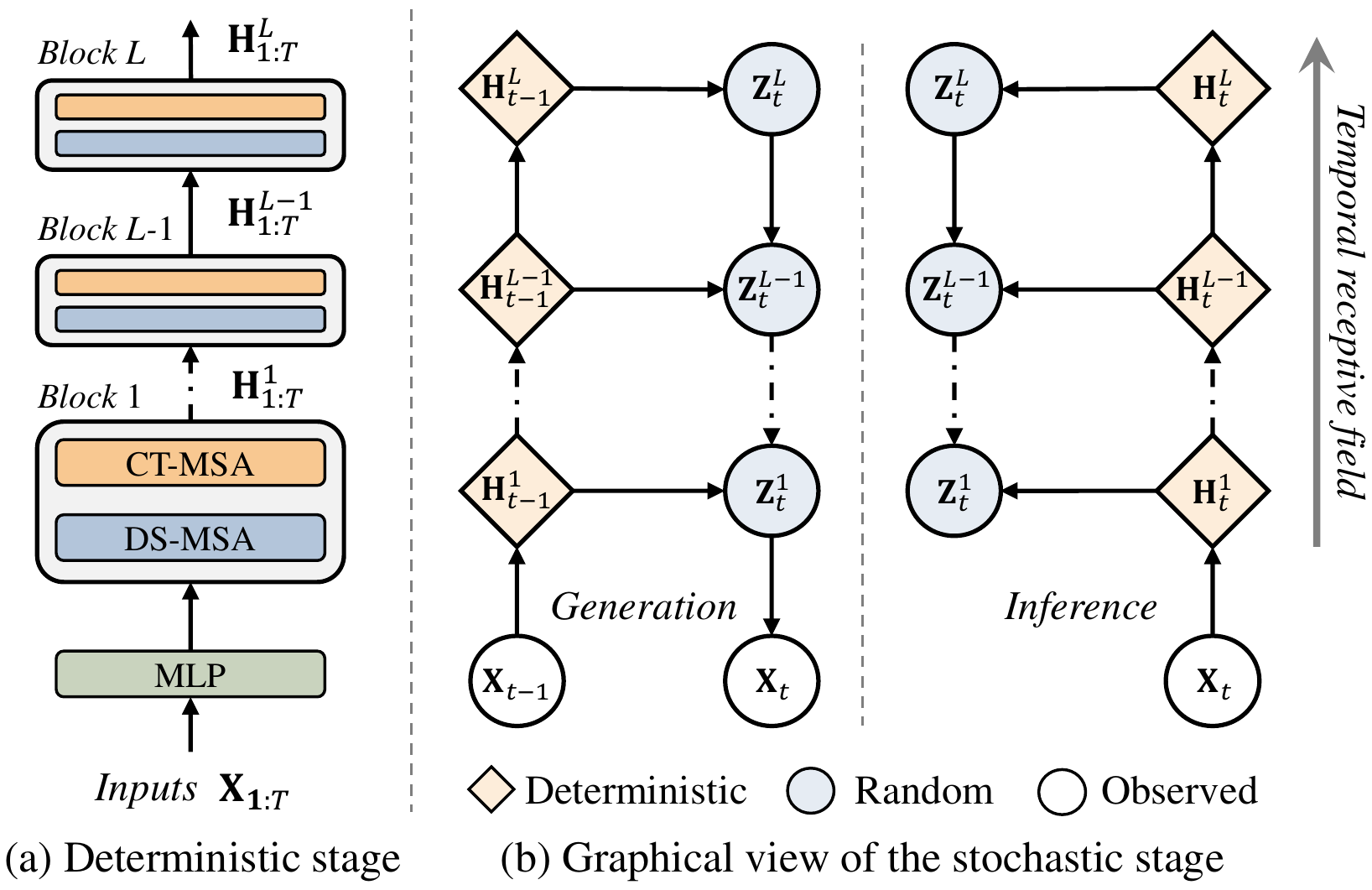}
  \vspace{-1.5em}
  \caption{The framework of AirFormer, which is separated into two stages: (a) One to leverage two types of MSA to learn deterministic spatio-temporal representations. (b) The other to capture the uncertainty via latent random variables. The temporal receptive field of the deterministic/stochastic states is gradually increased from bottom to top.}
  \label{fig:framework}
  
\end{figure}

Figure \ref{fig:framework} shows the framework of AirFormer for nationwide air quality prediction, which is decoupled into two stages:
\begin{itemize}[leftmargin=*]
    \item \emph{Bottom-up deterministic stage}: We first transform the historical readings $\mathbf{X}_{1:T}$ into the feature space using a multi-layer perceptron (MLP). The transformed features are then fed to $L$ AirFormer blocks to learn deterministic spatio-temporal representations. In each block, we factorize the space-time modeling along temporal and spatial domains, leading to the dual levels of MSA: DS-MSA for learning spatial interactions with linear complexity, and CT-MSA for capturing the temporal dependencies at each location. As shown in Figure \ref{fig:framework}(a), the output state at the $l$-th block is a 3D tensor, denoted as $\mathbf{H}^{l}_{1:T} \in \mathbb{R}^{T \times N \times C}$.
    
    \item \emph{Top-down stochastic stage}: Once the deterministic representations are obtained, we produce latent variables $\mathbf{Z}$ at each level. To maintain the parallelism of transformers, we adhere to \cite{sonderby2016ladder,aksan2018stcn} not establishing explicit dependencies between various time steps. Instead, we implicitly build temporal dependencies by conditioning a latent variable $\mathbf{Z}_t^{l-1}$ on its higher-level variable $\mathbf{Z}_t^{l}$ as shown in Figure \ref{fig:framework}(b), where $\mathbf{Z}_t^{i} \in \mathbb{R}^{N \times C}$ and $i=\{1,\dots,L\}$. In this way, the lower-level latent variables focus more on the local information, while the upper-level ones have a larger receptive field due to their corresponding deterministic input. In our model, the generation task is to predict the next time step given all past steps using the prior $p_\theta (\mathbf{Z}^l_t | \mathbf{X}_{1:t-1})$, and the inference task is to approximate the posterior $q_{\phi}\left(\mathbf{Z}_{t}^{l} | \mathbf{X}_{1:t} \right)$. As AirFormer belongs to the VAE family, we train our model by jointly optimizing the prediction loss and the ELBO. 
\end{itemize}

In the following parts, we start by introducing the modules for spatio-temporal representation learning in the deterministic stage, i.e., DS-MSA and CT-MSA. Next, we elaborate on the generation and inference model of the stochastic stage. Finally, we delineate the optimization of our model.

\subsection{Dartboard Spatial MSA (DS-MSA)}

\noindent \textbf{Motivation.} Besides local emissions, the air quality of a place is impacted by its neighbors as air pollutants are dispersed among various locations. Given the hidden state of $N$ stations at time $t$ ($\mathbf{H}_t \in \mathbb{R}^{N \times C}$), the complexity of standard MSA for spatial modeling is $\mathcal{O}(N^2 C)$, where $C$ is the feature length. This quadratic cost makes it inefficient to accommodate with fine-grained data, e.g., 1,000+ stations. 
\vspace{0.1em}

\begin{figure}[!h]
  \centering
  \vspace{-1em}
  \includegraphics[width=0.485\textwidth]{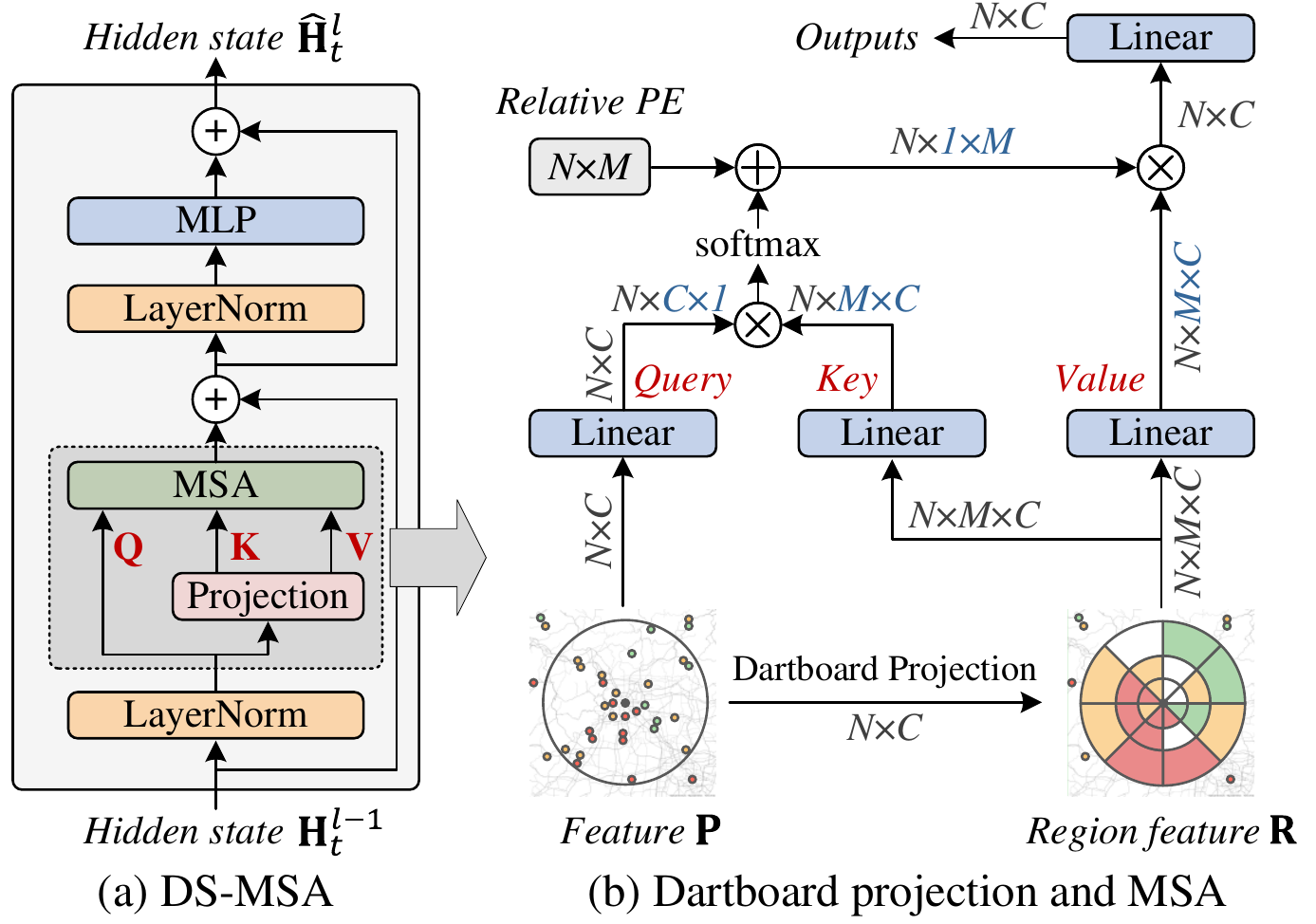}
  \vspace{-1.5em}
  \caption{Pipeline of DS-MSA. For simplicity, we illustrate the single-head version of MSA here. PE: position encoding.}
  \label{fig:ds_msa}
  \vspace{-0.6em}
\end{figure}

\noindent \textbf{Overview.} We present a novel MSA termed Dartboard Spatial MSA (DS-MSA) for \emph{efficiently} capturing the spatial dependencies among different locations at each time step. Notably, our DS-MSA possesses a large receptive field while only taking \emph{linear} complexity w.r.t the number of stations, which is inspired by the spatial feature extraction in \cite{zheng2015forecasting}. Figure \ref{fig:ds_msa}(a) presents the pipeline of DS-MSA. At the $l$-th block, DS-MSA takes the hidden state $\mathbf{H}^{l-1}_t$ as inputs. The features are first normalized and used to generate the queries with a linear layer. For each query station, we project its surroundings into the dartboard distribution to obtain the keys and values. As a result, the number of keys (or values) is reduced to $M$, where $M$ is the region number. We then perform MSA to learn spatial dependencies and finally utilize MLPs to produce the output $\hat{\mathbf{H}}_t^l \in \mathbb{R}^{N \times C}$.  

\noindent \textbf{Dartboard Projection \& MSA.} 
For a certain station $i$, we unify the formulation of the dartboard projection by introducing a \emph{projection matrix} $\mathbf{A}_i \in \mathbb{R}^{M \times N}$ which denotes how nearby stations are mapped to $M$ regions bound by the line fragments and circles, where each entry $a_{jk} \geq 0$ denotes the likelihood that the $k$-th station belongs to the region $j$. $\mathbf{A}_i$ has each row summing to zero and each non-zero entry in the same row is evenly distributed (like average pooling). 

Figure \ref{fig:dartboard} shows an example in which we partition the surroundings of a query station in a dartboard fashion. The outmost circle has the largest semidiameter (e.g. 500km), while the innermost one has the smallest (e.g., 50km). The three circles focus on a common center (the query station) and are further divided by four lines heading in various directions. The stations falling outside the outmost circle are not considered for MSA. In this case, we have $M=24+1=25$ ($+1$ for including the query station itself).

Assume the inputs of dartboard projection is $\mathbf{P} \in \mathbb{R}^{N \times C}$, we project the station features into regional representations that correspond to each station via the assignment matrix:
\begin{equation}
    \mathbf{R}_i = \mathbf{A}_i \mathbf{P}, \quad \mathbf{R} = \left[\mathbf{R}_1, \mathbf{R}_2, \cdots, \mathbf{R}_N \right],
\end{equation}
where $[\cdot,\cdot]$ is concatenation; $\mathbf{R} \in \mathbb{R}^{N \times M \times C}$ is the regional features after projection. Next, we generate the queries, keys and values using linear layers and perform MSA to capture the spatial relations between the query station and its corresponding regions (notations are detailed in Eq. \ref{eq:msa}):
\begin{equation}\label{eq:msa_with_bias}
    \mathbf{X}_{h}=\operatorname{Softmax}\left( \alpha \mathbf{Q}_{h} \mathbf{K}_{h}^{\top}  + \mathbf{B}_h \right) \mathbf{V}_{h},
\end{equation}
where $\mathbf{B}_h \in \mathbb{R}^{N \times M}$ is learnable relative position encoding \cite{wu2021rethinking} to incorporate position information. We can also encode the impacts of external factors (e.g., wind direction and speed) in $\mathbf{B}_h$ to improve performance. In particular, the regions without any stations are \emph{masked} during MSA. Figure \ref{fig:ds_msa}(b) can help better understand this process.

\vspace{0.2em}
\noindent \textbf{Discussion.} DS-MSA is designed by jointly considering the following factors: 1) \emph{Spatial dependencies}: Considering the domain knowledge of air pollution dispersion that the spatial correlations between nearby locations are always stronger than those far away, each station attends to its surroundings in a dartboard fashion, i.e., close places at a fine granularity and the distant regions at a coarse granularity. 2) \emph{Efficiency}: Since the number of regions $M$ is small ($M \ll N$), the computational complexity $\mathcal{O}(MNC)$ grows linearly with the increase of the number of stations, which is more efficient than standard MSA. 3) \emph{Lightweight}: By using the dartboard projection, DS-MSA does not introduce extra learnable parameters into standard MSA, thus being lightweight in practice.

\subsection{Causal Temporal MSA (CT-MSA)}
\begin{figure}[!t]
  \centering
  \includegraphics[width=0.47\textwidth]{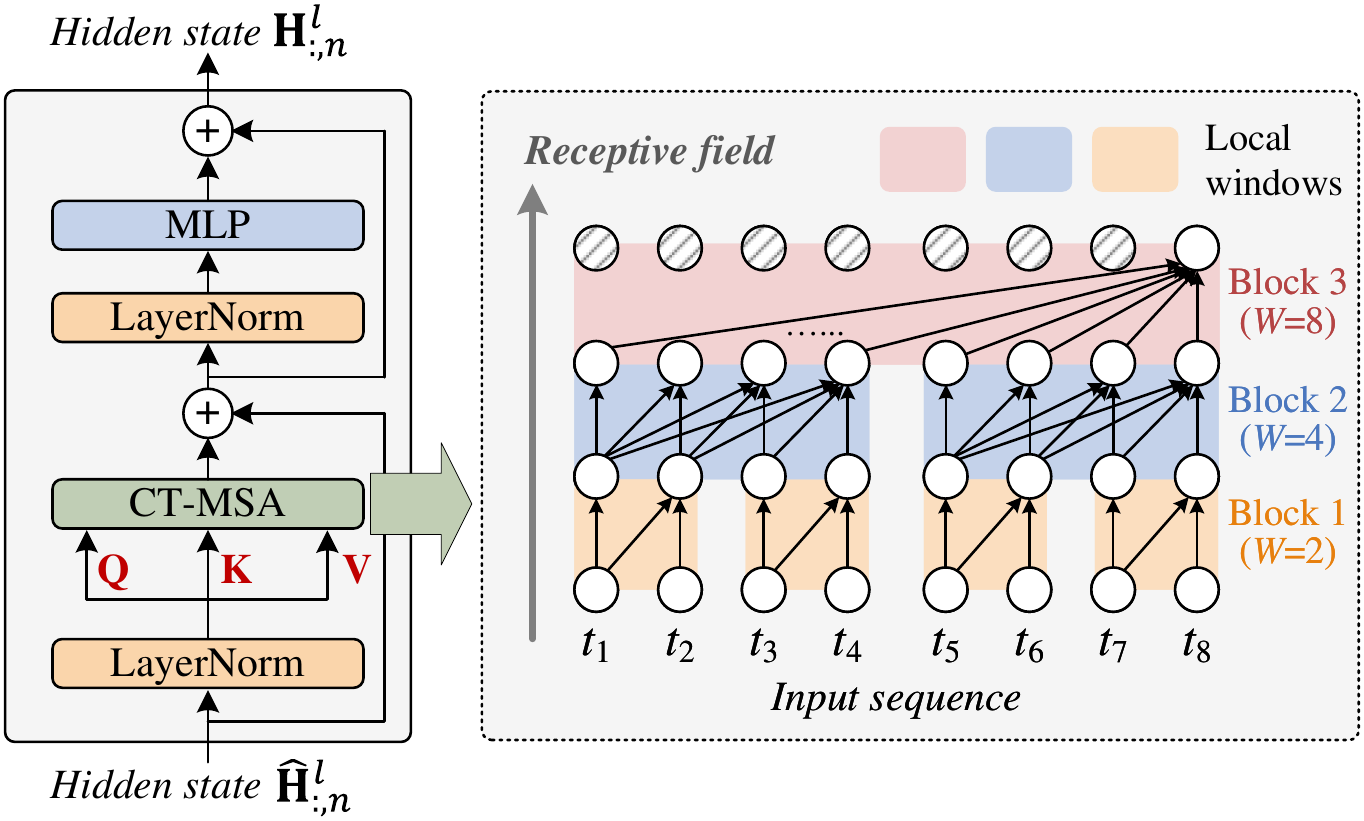}
  \vspace{-1.5em}
  \caption{Left: CT-MSA. Right: We use an example of 8 time steps for illustration. Here, CT-MSA at each AirFormer block has local windows of size 2, 4 and 8, respectively.}
  \label{fig:ct_msa}
  \vspace{-1.0em}
\end{figure}

In addition to spatial dependencies, the air quality of a location depends on its history. Given the hidden state of a location across all past steps $\hat{\mathbf{H}}^{l}_{:,n} \in \mathbb{R}^{T \times C}$ (the output of DS-MSA), adopting MSA to learn temporal dependencies induces quadratic cost w.r.t the number of time steps $T$. Here, we propose Causal Temporal MSA (CT-MSA) as a strong and efficient alternative to standard MSA for temporal modeling. As depicted in Figure \ref{fig:ct_msa}, its pipeline is mostly identical to standard MSA. Leveraging the domain knowledge of time series, CT-MSA has two major modifications:

\vspace{0.2em}
\noindent \textbf{Local window.} Considering that nearby time steps usually have stronger correlations than faraway slots, we perform MSA within non-overlapping windows to capture local interactions among time steps, leading to $\mathcal{O} (T W C)$ cost ($\frac{T}{W}$ times less than standard MSA), where $W$ is the window size. To preserve the large receptive field of standard MSA, we gradually increase the window size in different AirFormer blocks (from bottom to top), as shown in Figure \ref{fig:ct_msa}.

\vspace{0.2em}
\noindent \textbf{Temporal causality.} Since the air quality at the current step is not conditioned on its future, we follow WaveNet \cite{oord2016wavenet} to introduce causality into MSA (see Figure \ref{fig:ct_msa}), which ensures the model cannot violate the temporal order of input data. Such causality can be easily implemented by masking the specific entries in the attention map. To allow position-aware MSA, we add learnable absolute position encoding \cite{vaswani2017attention} to the input of CT-MSA.

\subsection{Top-Down Stochastic Stage}
After obtaining the deterministic representations, we build latent random variables to learn the \emph{uncertainty} of air quality data, e.g., unpredictable factors and noisy observations. 

\vspace{0.1em}
\noindent \textbf{Generation Model.} As shown in Figure \ref{fig:framework}(b), the generation model aims to predict the next step given all \emph{past} steps. As we have encoded the spatial dependencies among locations in deterministic states $\mathbf{H}_t$, we can factorize the prior distribution of a set of random variables $\mathcal{Z}_t = \{\mathbf{Z}_t^1,\dots,\mathbf{Z}_t^L\}$ as:
\begin{equation}\label{eq:generation}
\begin{split}
    &p_{\theta}\left(\mathcal{Z}_{t} | \mathbf{X}_{1:t-1}\right) = \prod_{n=1}^{N} p_{\theta}\left(\left\{\mathbf{z}_{t,n}^1,\dots,\mathbf{z}_{t,n}^L \right\}| \mathbf{X}_{1:t-1}\right) 
    \\&=\prod_{n=1}^{N} p_{\theta}\left(\mathbf{z}_{t,n}^{L} | \mathbf{h}_{t-1,n}^{L}\right) \prod_{l=1}^{L-1} p_{\theta}\left(\mathbf{z}_{t,n}^{l} | \mathbf{z}_{t,n}^{l+1}, \mathbf{h}_{t-1,n}^{l}\right),
\end{split}
\end{equation}
where $\mathbf{z}_{t,n}^l \in \mathbb{R}^C$ and $\mathbf{h}_{t,n}^l \in \mathbb{R}^C$ is the $n$-th row of $\mathbf{Z}_t^l$ and $\mathbf{H}_t^l$, respectively. In Eq. (\ref{eq:generation}), we follow VAE to set the prior distribution at each layer as a Gaussian distribution, i.e.,
\begin{equation}
p_{\theta}\left(\mathbf{z}_{t,n}^{l} | \mathbf{z}_{t,n}^{l+1}, \mathbf{h}_{t-1,n}^{l}\right)=\mathcal{N}\left(\mu_{t}^{l}, \sigma_{t}^{l}\right),    
\end{equation}
where the mean $\mu_{t}^{l}$ and the diagonal covariance $\sigma_{t}^{l}$ are parameterized by a neural network $f^l(\mathbf{z}_{t,n}^{l+1}, \mathbf{h}_{t-1,n}^{l})$ shared by all locations. In this way, we implicitly build connections between hidden states at different time steps, as the upper-layer random variables (e.g., $\mathbf{z}_{t,n}^{l+1}$) contain more contextual information due to their larger receptive field.

\vspace{0.1em}
\noindent \textbf{Inference Model.} In contrast, the inference model (see Figure \ref{fig:framework}(b)) is applied to approximate the posterior distribution of $\mathcal{Z}_t$ given both the \emph{current} and \emph{previous} steps:
\begin{equation}
\vspace{-0.2em}
\nonumber
    q_{\phi}\left(\mathcal{Z}_{t} | \mathbf{X}_{1:t}\right) =\prod_{n=1}^{N} q_{\phi}\left(\mathbf{z}_{t,n}^{L} | \mathbf{h}_{t,n}^{L}\right) \prod_{l=1}^{L-1} q_{\phi}\left(\mathbf{z}_{t,n}^{l} | \mathbf{z}_{t,n}^{l+1}, \mathbf{h}_{t,n}^{l}\right),
\end{equation}
\begin{equation}\label{eq:inference}
    \text{where} \quad q_{\phi}\left(\mathbf{z}_{t,n}^{l} | \mathbf{z}_{t,n}^{l+1}, \mathbf{h}_{t,n}^{l}\right)=\mathcal{N}\left(\hat{\mu}_{t}^{l}, \hat{\sigma}_{t}^{l}\right).
\end{equation}
Eq. (\ref{eq:inference}) employs the same factorization as Eq. (\ref{eq:generation}). The arguments of the Gaussian distribution at each layer is parameterized by a neural network $g^l (\mathbf{z}_{t,n}^{l+1}, \mathbf{h}_{t,n}^{l})$. Following a similar procedure as the generation model, the random variables generated by the posterior distribution can also effectively consider the spatio-temporal dependencies within air quality, thereby enhancing the predictive performance.

\subsection{Prediction \& Optimization}
AirFormer makes predictions based on both the deterministic and stochastic hidden states at time $T$ using an MLP:
\begin{equation}
    \hat{\mathbf{Y}}_{1:\tau} = \operatorname{MLP} ([\mathbf{H}_T^1, \dots, \mathbf{H}_T^L, \mathbf{Z}_T^1, \dots, \mathbf{Z}_T^L]).
\end{equation}
To train our model, we jointly optimize two loss functions:
\begin{equation}\label{eq:loss}
    \mathcal{L}=\mathcal{L}_{pred} + \mathcal{L}_{\text{ELBO}}.
\end{equation}
$\mathcal{L}_{pred}$ denotes the L1 loss between the prediction $\hat{\mathbf{Y}}_{1:\tau}$ and the corresponding ground truth $\mathbf{Y}_{1:\tau}$; the second term $\mathcal{L}_{\text{ELBO}}$ indicates the sum of \emph{negative} ELBO at all historical steps. Concretely, $\mathcal{L}_{\text{ELBO}}$ at each step is computed as:
\begin{equation}\label{eq:elbo}
\begin{aligned}
\begin{aligned}
\mathcal{L}_{\text{ELBO}}&(\theta, \phi;t) = \mathcal{L}_{rec}(\theta, \phi;t) + \mathcal{L}_{kl}(\theta, \phi;t) \\
&=-\mathbb{E}_{q_{\phi}\left(\mathcal{Z}_{t} | \mathbf{X}_{t}\right)}\left[\log p_{\theta}\left(\mathbf{X}_t | \mathcal{Z}_{t} \right)\right] 
\\ & \quad + KL\left(q_{\phi}\left(\mathcal{Z}_{t} | \mathbf{X}_{1: t}\right) \| p_{\theta}\left(\mathcal{Z}_{t} | \mathbf{X}_{1: t-1}\right)\right),
\end{aligned}
\end{aligned}
\end{equation}
where the first term is the reconstruction likelihood and the second term is the KL divergence between the prior and the posterior; $\mathcal{Z}_t = \{\mathbf{Z}_t^1,\dots,\mathbf{Z}_t^L\}$ is a set of latent variables at each layer. We further simplify the KL term ($\mathcal{L}_{kl}$) using the factorization in Eq. (\ref{eq:generation}) and Eq. (\ref{eq:inference}), resulting in 
\begin{equation}
\nonumber
\begin{aligned}
\mathcal{L}_{kl} &= \sum_{n=1}^{N}  KL \left(q_{\phi}\left(\mathbf{z}_{t,n}^{L} | \mathbf{h}_{t,n}^{L}\right) \| p_{\theta}\left(\mathbf{z}_{t,n}^{L} | \mathbf{h}_{t-1,n}^{L}\right)\right) + \sum_{n=1}^{N} \sum_{l=1}^{L-1} \\ 
&  \mathbb{E}_{q_{\phi}}\left[KL \left(q_{\phi}\left(\mathbf{z}_{t,n}^{l} | \mathbf{z}_{t,n}^{l+1}, \mathbf{h}_{t,n}^{l}\right) \| p_{\theta}\left(\mathbf{z}_{t,n}^{l} | \mathbf{z}_{t,n}^{l+1}, \mathbf{h}_{t-1,n}^{l}\right)\right)\right]
\end{aligned}
\end{equation}

\section{Experiments}
\subsection{Dataset Description}
Our system collected nationwide air quality data and meteorological data for a four-year period (Jan. 1, 2015 to Dec. 31, 2018) throughout the Chinese mainland. The system details can be found in \cite{yi2018deep}.
\begin{itemize}[leftmargin=*]
    \item \emph{Air quality data}: We collected hourly air quality data from 1,976 stations covering 342 cities in China. Each air quality record contains the concentration of six types of pollutants, including PM$_{2.5}$, PM$_{10}$, NO$_{2}$, CO, O$_{3}$, and SO$_{2}$. Among them, the primary pollutant of air quality is PM$_{2.5}$, thus we employ its reading as the \emph{prediction target}.
    \item \emph{Meteorological data}: This dataset is collected every hour from 2,575 monitoring stations in China. Each record consists of weather conditions (e.g., sunny, rainy), temperature, humidity, wind speed, and wind direction. 
\end{itemize}

We choose the air quality monitoring stations with a missing rate of PM$_{2.5}$ less than 20\% to conduct our experiments, leading to a 1,085-station dataset. Figure \ref{fig:intro} illustrates the station distribution. Our dataset has extensive coverage in both space and time scales compared to previous datasets (see Table \ref{tab:addlabel}). Following the prior studies \cite{liang2018geoman,wang2020pm25}, we predict the concentration of PM$_{2.5}$ over the next 24 steps given the past 24 steps, where each step stands for 3 hours. In other words, we perform a 72-hour future prediction based on the readings from the past 72 hours. The dataset is partitioned in chronological order with the first two years for training, the third year for validation, and the last year for testing. Z-score normalization is applied to the model inputs for fast convergence. More dataset details can be found in the Data Appendix.

\begin{table}[!t]
  \small
  \centering
  \tabcolsep=1.2mm
   \caption{Dataset comparison. m: month. Joint means collectively predicting the air quality at all locations.}
  \vspace{-0.5em}
    \begin{tabular}{l|ccccc}
    \shline
    Dataset & Venue & \#Node & Range & Scale & Joint? \\
    \hline
    \cite{zheng2015forecasting} & KDD & 35 & 48m & City & \xmark \\
    \cite{yi2018deep} & KDD & 35 & 48m & City & \xmark \\
    \cite{lin2018exploiting}  & GIS & 35 & 14m & City & \checkmark \\
    \cite{wang2020pm25} & GIS & 184   & 48m & Nation & \checkmark \\
    \cite{xu2021highair} & preprint & 56 & 12m & Province & \checkmark \\
    \cite{chen2021group} & preprint & 209 & 28m & Nation & \checkmark \\
    \rowcolor{background_gray}  
    Our dataset  & N/A & \textbf{1,085}  & 48m & Nation & \checkmark \\
    \shline
    \end{tabular}%
  \label{tab:addlabel}%
  \vspace{-1em}
\end{table}%



\subsection{Implementation Details}
We implement our model by PyTorch 1.10 using a Quadro RTX A6000 GPU. The Adam optimizer \cite{kingma2014adam} is utilized to train our model, and the batch size is 16. The learning rate starts from $5 \times 10^{-4}$, halved every three epochs. The number of AirFormer blocks is set to 4 with local window size $W=3,6,12,24$, respectively. For the hidden dimension $C$ in DS-MSA and CT-MSA, we conduct a grid search over $\{8, 16, 32, 64\}$, and $C=32$ obtains the best result. The number of heads in DS-MSA and CT-MSA is 2. In DS-MSA, we partition the space into regions by two circles with 50 and 200km semidiameter, respectively. $f$ and $g$ in the stochastic stage are parameterized by 3-layer MLPs. We will detail the effects of these hyperparameters later.

\subsection{Baselines for Comparison}
We compare our AirFormer with the following baselines that belong to the following four categories:
\begin{itemize}[leftmargin=*]
    \item \emph{Classical methods}: \textbf{HA} is history average which predicts air quality by the average value of historical readings in the corresponding periods. \textbf{VAR} is vector autoregression.
    \item \emph{STGNN variants}: we also take some popular STGNNs as baselines, i.e., \textbf{DCRNN} \cite{li2017diffusion}, \textbf{STGCN} \cite{yu2018spatio}, \textbf{GWNET} \cite{wu2019graph}, and \textbf{MTGNN} \cite{wu2020connecting}. They generalize well to our application.
    \item \emph{Attention-based models}: \textbf{ASTGCN} \cite{guo2019attention} and \textbf{GMAN} \cite{zheng2020gman} are attention-based baselines for spatio-temporal forecasting. \textbf{STTN} \cite{xu2020spatial} is a variant of transformers for traffic forecasting, which can be easily adapted to air quality prediction.
    \item \emph{Air quality prediction}: we select three strong models for a comparison, including \textbf{DeepAir} \cite{yi2018deep}, \textbf{PM$_{2.5}$-GNN} \cite{wang2020pm25} and \textbf{GAGNN} \cite{chen2021group}. Some \cite{pan2018hyperst,wang2021modeling,wang2022air,liang2022mixed} are omitted due to out-of-GPU-memory (OOM).
\end{itemize}

For a fair comparison, we tune different hyperparameters for all baselines, finding the best setting for each. See more details about their settings in the Technical Appendix.

\subsection{Evaluation Metrics}
We leverage Mean Absolute Error (MAE) and Root Mean Squared Error (RMSE) for evaluation, where a smaller metric means better performance. Moreover, we follow \cite{zheng2015forecasting,yi2018deep} to discuss the errors on predicting sudden changes. The sudden changes are defined as the cases where PM$_{2.5}$ is larger than 75 µg/m$^3$ and changes more than a threshold (i.e., $\pm$20 µg/m$^3$) in the next three hours.

\begin{table*}[!t]
  \centering
  \small
  \tabcolsep=2.3mm
  \caption{5-run results. The magnitude of \#Param (the number of parameters) is Kilo. The bold/underlined font mean the best/the second best result. * denotes the improvement over the second best model is statistically significant at level 0.01 \protect \cite{kim2015t}.}
  \vspace{-0.5em}
    \begin{tabular}{l||c|cc|cc|cc|cc}
    \shline
    \multirow{2}*{\textbf{Model}} & \multicolumn{1}{c|}{\multirow{2}*{\textbf{\#Param (K)}}} & \multicolumn{2}{c|}{\textbf{1-24h}} & \multicolumn{2}{c|}{\textbf{25-48h}} & \multicolumn{2}{c|}{\textbf{49-72h}} & \multicolumn{2}{c}{\textbf{Sudden change}} \\
\cline{3-10}          &       & \textbf{MAE} & \textbf{RMSE} & \textbf{MAE} & \textbf{RMSE} & \textbf{MAE} & \textbf{RMSE} & \textbf{MAE} & \textbf{RMSE} \\
    \hline
    \hline
    HA \cite{zhang2017deep}  & -     & 31.25 & 62.52 & 31.19 & 62.49 & 31.16 & 62.49 & 72.89 & 132.42 \\
    VAR \cite{toda1991vector} & -     & 29.91 & 52.10 & 31.61 & 64.59 & 30.18 & 65.10 & 70.86 & 114.78 \\
    \hline
    DCRNN \cite{li2017diffusion}  & 397   & 19.35 & 46.40 & 24.06 & 57.38 & 25.30 & 58.24 & 63.22 & 112.30 \\
    STGCN \cite{yu2018spatio}  & 453   & 19.06 & 42.69 & 24.09 & 56.50 & 25.10 & 58.96 & 61.35 & 111.89 \\
    GWNET  \cite{wu2019graph} & 825   & 17.79 & 39.49 & \underline{23.32} & 53.17 & 25.00 & 57.01 & 59.33 & 105.75 \\
    MTGNN \cite{wu2020connecting} & 207   & 18.15 & 38.99 & 23.47 & \underline{52.21} & \underline{24.77} & \underline{55.73} & \underline{59.24} & \underline{103.71} \\
    \hline
    ASTGCN  \cite{guo2019attention}  & 4,812  & 20.76 & 50.29 & 24.37 & 56.04 & 25.22 & 57.77 & 63.19 & 114.57 \\
    GMAN \cite{zheng2020gman} & 269   & 19.60 & 45.70 & 23.79 & 54.25 & 24.89 & 56.33 & 61.83 & 109.57 \\
    STTN \cite{xu2020spatial} & 188   & 18.22 & \underline{37.44} & 24.16 & 52.91 & 25.35 & 56.14 & 60.36 & 105.20 \\
    \hline
    DeepAir \cite{yi2018deep} & 183   & \underline{17.47} & 39.12 & 23.40 & 53.48 & 24.95 & 56.92 & 60.26 & 109.95 \\
    PM$_{2.5}$-GNN \cite{wang2020pm25} & 101   & 20.20 & 48.96 & 25.04 & 59.56 & 26.31 & 60.46 & 63.64 & 119.00 \\
    GAGNN \cite{chen2021group} &  412  & 
    19.53 & 45.68 & 24.56 & 58.59 & 25.56 & 59.61 & 64.38 & 116.51 \\
    \rowcolor{background_gray}
        AirFormer (ours) & 246 & \textbf{16.03}* & \textbf{32.36}* & \textbf{21.65}* & \textbf{44.67}* & \textbf{23.64}* & \textbf{50.22}* &  \textbf{54.92}* &  \textbf{90.15}* \\
    \shline
    \end{tabular}%
  \label{tab:results}%
  \vspace{-1em}
\end{table*}%

\subsection{Model Comparison}
\vspace{-0.1em}
In this section, we perform a model comparison in terms of MAE and RMSE. We run each method five times and report the average metric of each model. As shown in Table \ref{tab:results}, AirFormer \emph{significantly} outperforms all competing baselines on both metrics according to the Student's T-test \cite{kim2015t} at level 0.01. In contrast to the second best method\footnote{The model with the second least mean MAE over all horizons.} (i.e., DeepAir), AirFormer reduces the MAE by 8.2\%, 7.5\% and 5.3\% on 24-, 48- and 72-hour future prediction, respectively. When predicting the sudden changes, AirFormer achieves at least 7.3\% and 13.1\% improvements on MAE and RMSE thanks to the robustness provided by stochastic latent spaces. 

From Table \ref{tab:results}, we can also observe that: 1) Deep-learning-based approaches surpass the classical methods (HA and VAR) by a large margin due to their greater learning capacity. 2) Although STGNN-based models, e.g., GWNET and MTGNN, are originally evaluated on traffic forecasting, they also generalize well on air quality prediction, showing comparable results to the second best method DeepAir. 3) Our AirFormer clearly outperforms STTN which is also a transformer model. This verifies that the domain knowledge of air pollution not only helps us design our model with more interpretations, but also enhances the predictive accuracy.

\vspace{-0.3em}
\subsection{Ablation Study}
\vspace{-0.1em}
\noindent \textbf{Effects of DS-MSA.} To study the effects of DS-MSA, we consider the following variants for comparison: a) \textbf{w/o DS-MSA}: We turn off DS-MSA in AirFormer, i.e., no spatial modeling in this model. b) \textbf{MSA}: We replace DS-MSA with standard MSA. c) \textbf{Local MSA}: DS-MSA is substituted by local MSA in which each query station attends to its neighbors within 500km. The results are shown in the upper part of Table \ref{tab:ds_msa}. First, we find removing DS-MSA leads to a significant degradation on MAE, revealing the great importance of addressing spatial dependencies. Second, DS-MSA obtains lower prediction errors over all future horizons, while running \textbf{39\%} faster than vanilla MSA and 21\% faster than local MSA. This merit suggests that our DS-MSA has great potential to be a basic building block for capturing spatial dependencies within air quality data in practice.

We then discuss different settings of the dartboard. In the lower part of Table \ref{tab:ds_msa}, $r_1$-$r_2$-$r_{3}$ means we divide the space by 3 circles with $r_1$, $r_2$, and $r_3$ km semidiameter. 50-200 denotes a 2-circle partition with 50 and 200km semidiameter. Compared to the 1-circle partition with 50km semidiameter, the 50-200 dartboard achieves lower errors due to a larger receptive field. Despite the lowest MAE of 50-200-500, it inevitably results in more computational costs (33.8\% slower than 50-200). To make a better trade-off between speed and accuracy, we choose 50-200 as our default setting.
\begin{table}[!h]
  \small
  \centering
  \tabcolsep=1.3mm
  \vspace{-1em}
  \caption{Effects of DS-MSA over MAE. The gray row is our default setting. Time/epo: seconds per training epoch. }
  \vspace{-0.5em}
    \begin{tabular}{l||c|ccc}
    \shline
    \textbf{Variant} & \textbf{Time/epo} & \textbf{1-24h} & \textbf{25-48h} & \textbf{49-72h}  \\
    \hline\hline
    w/o DS-MSA &  1,284  & 17.24 & 23.55  & 24.96  \\
    MSA  & 2,818 & 18.19 & 23.24 & 24.23 \\
    Local MSA & 2,157 & 18.63 & 23.79 & 24.98  \\
    \rowcolor{background_gray}
    DS-MSA (50-200) &  1,708  & \textbf{16.03} & \underline{21.65}  & \underline{23.64}  \\
    \hline
    DS-MSA (50)    &  1,547  & 16.43 & 22.87 & 24.32  \\
    DS-MSA (50-200-500) & 2,285 & \underline{16.09} & \textbf{21.30} & \textbf{22.85}   \\
    \shline
    \end{tabular}%
  \label{tab:ds_msa}%
  \vspace{-0.5em}
\end{table}%

\begin{figure}[!h]
  \centering
  \vspace{-1em}
  \includegraphics[width=0.48\textwidth]{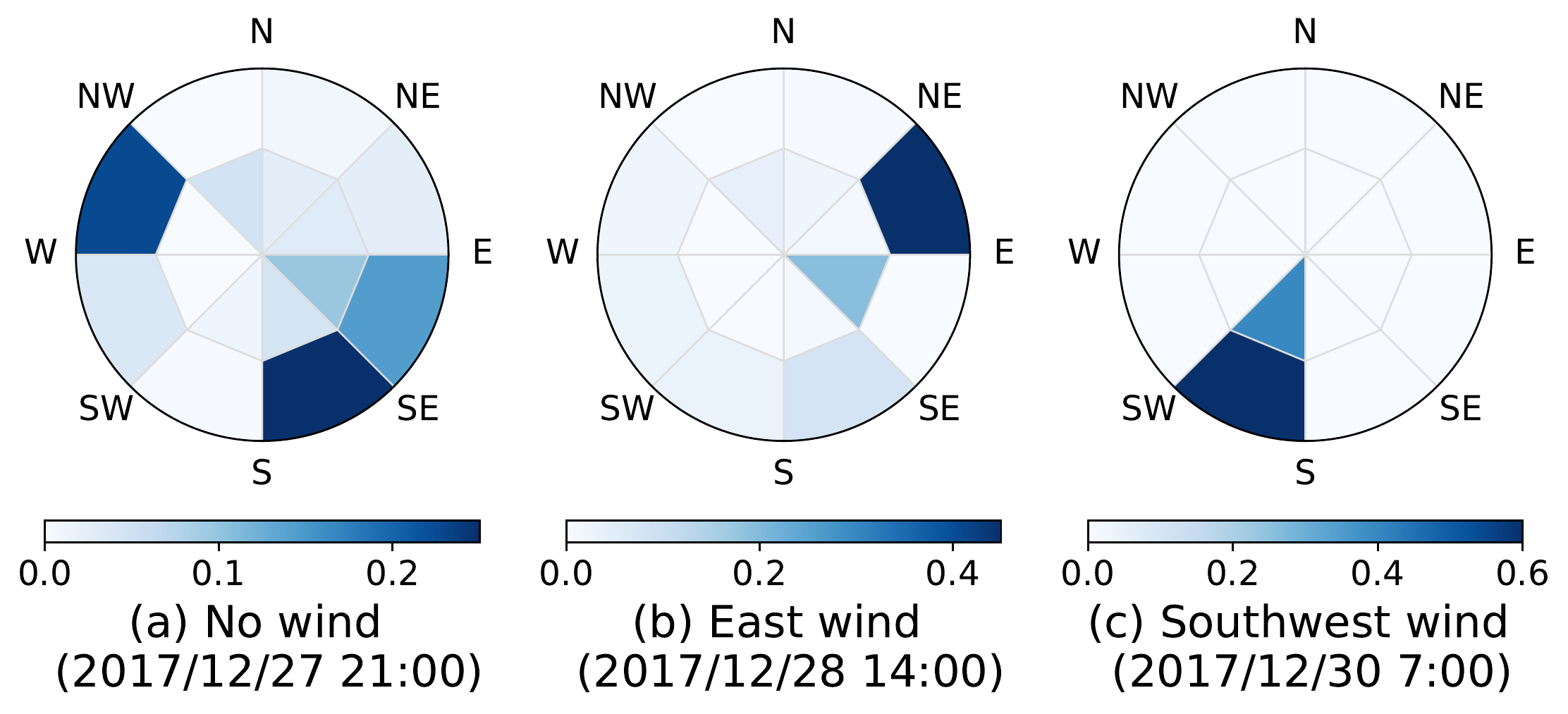}
  \vspace{-1.5em}
  \caption{Visualization of DS-MSA at the first block. We omit the attention weight at the center station (query) itself.}
  \label{fig:vis_dartboard}
\end{figure}

To further investigate DS-MSA, we perform a case study on the 50-200 dartboard centered by Xizhimen (a hybrid district in Beijing) from Dec. 27--30, 2017. In Figure \ref{fig:vis_dartboard}, the attention weights are dispersed (almost $<$0.2) when there is no wind. If the wind comes from the east or the southwest, the attention weights accordingly become more concentrated on the corresponding directions. In light of this study, DS-MSA is not only effective but can also be easily interpreted.

\begin{figure}[!h]
  \centering
  \vspace{-1em}
  \includegraphics[width=0.48\textwidth]{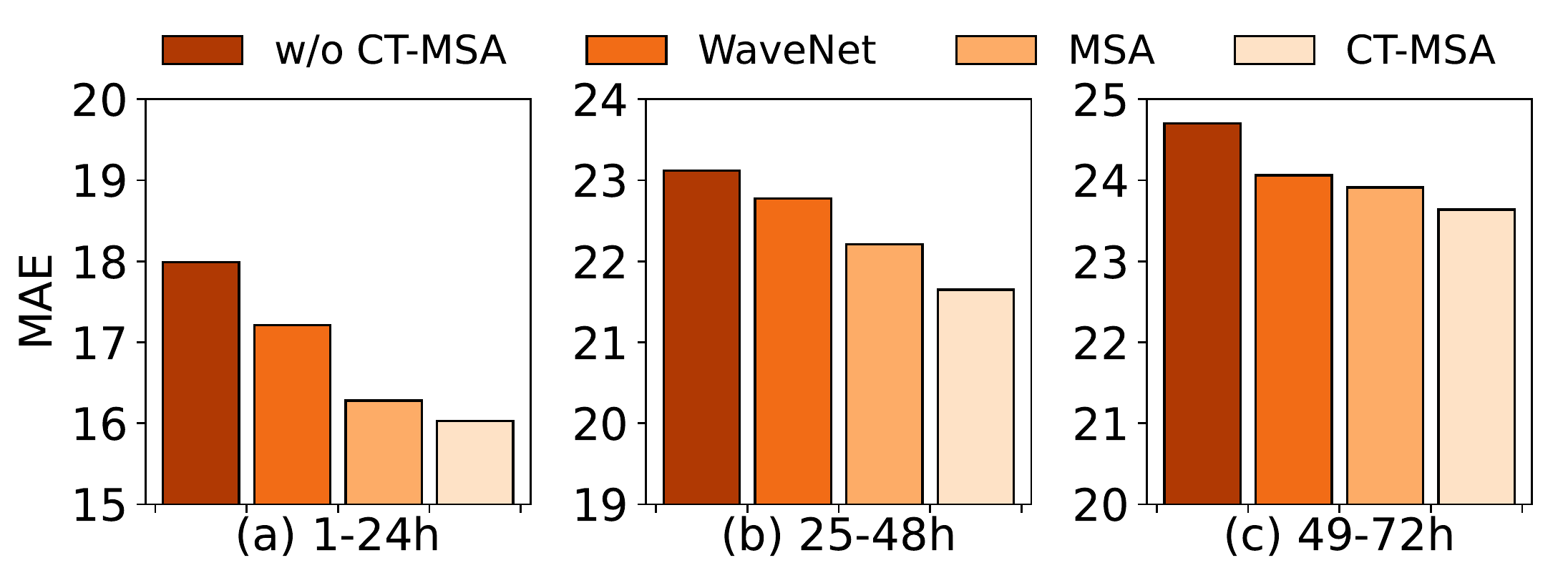}
  \vspace{-1.5em}
  \caption{Effects of CT-MSA on MAE.}
  \label{fig:exp_ct_msa}
  \vspace{-0.5em}
\end{figure}

\noindent \textbf{Effects of CT-MSA.}
To examine the efficacy of CT-MSA for capturing temporal dependencies, we compare our model with its variants integrated with various temporal modules: a) \textbf{w/o CT-MSA}: we remove CT-MSA from our AirFormer. b) \textbf{MSA}, \textbf{WaveNet}: we replace CT-MSA by standard MSA or WaveNet \cite{wu2019graph}. The results are shown in Figure \ref{fig:exp_ct_msa}. Primarily, all the variants with temporal modules perform much better than w/o CT-MSA, verifying the necessity of temporal modeling. Moreover, both MSA-based methods surpass WaveNet, which reveals the superiority of MSA in air quality modeling. Remarkably, integrating causality and local windows into MSA consistently improves the performance over all future steps (see MSA vs. CT-MSA).

\noindent \textbf{Effects of Latent Variables.} As a crucial component of AirFormer, the stochastic stage empowers our model to capture the uncertainty within air quality data, thus boosting the performance. To investigate its effectiveness, we compare our AirFormer with its variant that turns off the stochastic stage. As shown in Table \ref{tab:stochastic}, integrating latent variables reduces the MAE on sudden changes by 4.5\% while introducing little additional time (149 seconds) per training epoch. 

\begin{table}[!h]
  \centering
  \small
  \tabcolsep=3mm
  \vspace{-0.5em}
  \caption{Effects of latent variables on MAE. w/o: without.}
  \vspace{-0.5em}
    \begin{tabular}{l|c|c|c}
    \shline
    \textbf{Varaint} & \textbf{Time/epo} & \textbf{1-72h} & \textbf{Sudden change} \\
    \hline \hline
    w/o Stochastic & 1,559 & 20.97 & 57.52 \\
    \rowcolor{background_gray}
    AirFormer & 1,708 & \textbf{20.44} & \textbf{54.92} \\
    \shline
    \end{tabular}%
  \label{tab:stochastic}%
  \vspace{-0.5em}
\end{table}%

\begin{figure}[!h]
\vspace{-0.5em}
  \centering
  \includegraphics[width=0.47\textwidth]{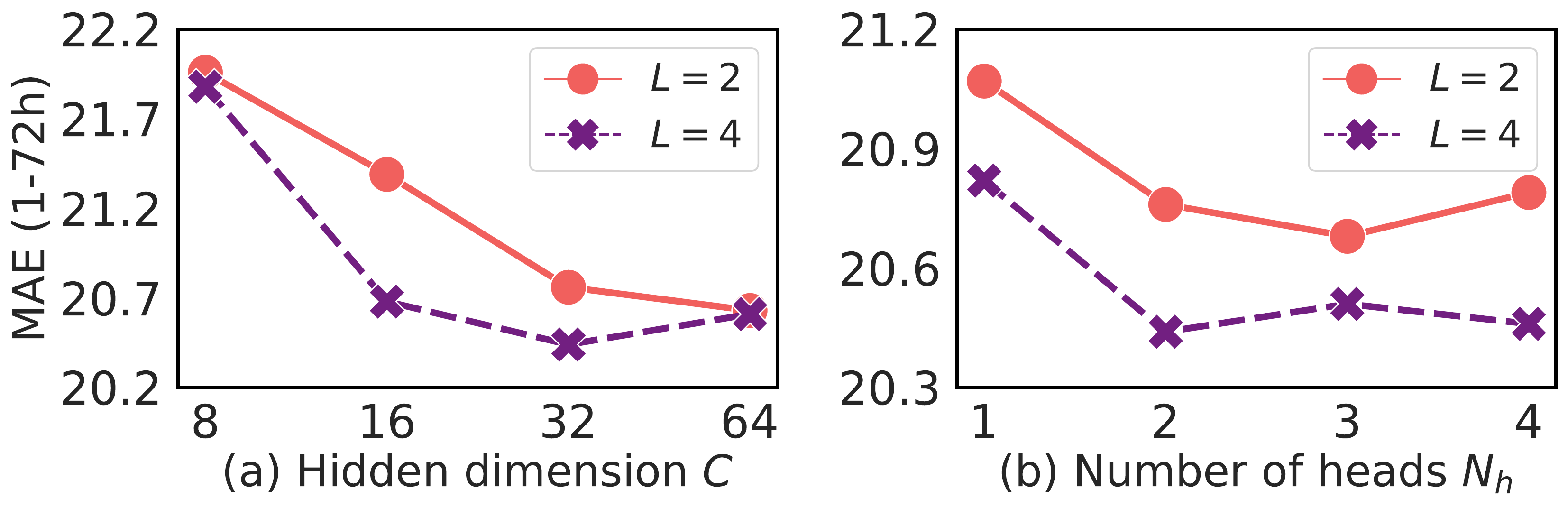}
  \vspace{-1.5em}
 \caption{Effects of hyperparameters on MAE.} 
  \label{fig:exp_hyperparam}
\end{figure}

\noindent \textbf{Hyperparameter Study.}
Next, we investigate the effects of the number of AirFormer blocks $L$, the hidden dimension $C$, and the number of heads $N_h$. Figure \ref{fig:exp_hyperparam}(a) shows the results of AirFormer with 2 or 4 blocks vs. different $C$, from which we observe: 1) Stacking more AirFormer blocks consistently achieves lower MAE. We encounter the OOM issue when $L>4$ and thus set $L=4$ as our default setting. 2) Using a very small $C$ (e.g., 16) leads to worse performance due to its limited capacity. 3) $C=64$ obtain similar accuracy. In Figure \ref{fig:exp_hyperparam}(b), we find that the performance of AirFormer is not sensitive to the number of heads when $N_h>1$.

\begin{figure}[!h]
  \centering
  \vspace{-1em}
  \includegraphics[width=0.48\textwidth]{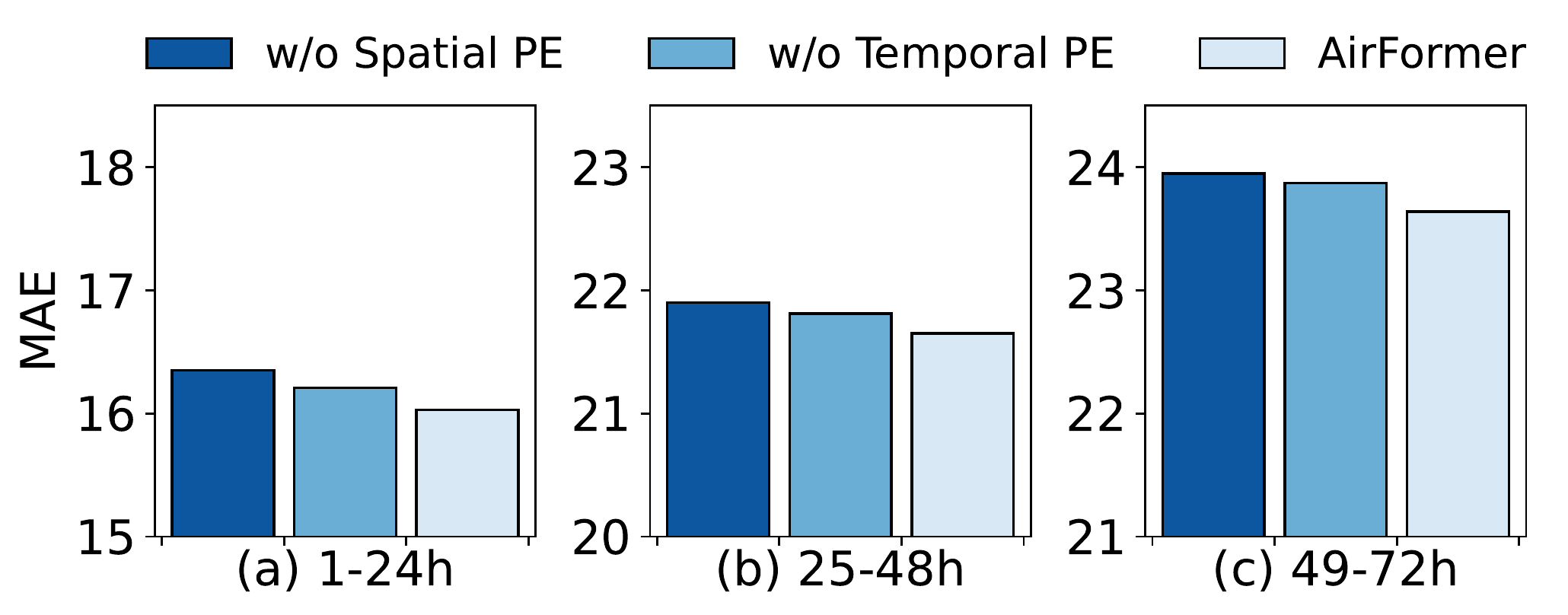}
  \vspace{-1.5em}
  \caption{Effects of position encoding (PE) on MAE.}
  \label{fig:exp_pe}
  \vspace{-1.2em}
\end{figure}

\noindent \textbf{Effects of Position Encoding.} Since MSA is permutation-invariant, we integrate position encoding (PE) into DS-MSA and CT-MSA to consider the order information. As depicted in Figure \ref{fig:exp_pe}, removing either Spatial PE in DS-MSA or Temporal PE in CT-MSA will cause degenerated performance across all future horizons. Besides, the improvement of integrating Spatial PE is slightly higher than Temporal PE.

\section{Related Works}
Accurately forecasting air quality is one of the core tasks in smart city efforts with significant social impacts, 

\noindent \textbf{Physics-based models} represent the emission and dispersion of air pollutants as a dynamic system and simulate the process using numerical functions. They investigate the primary causes of air pollution, including chemicals, vehicles, and factories \cite{vardoulakis2003modelling,arystanbekova2004application,daly2007air}. However, it is non-trivial to acquire such a variety of data sources precisely. 

\noindent \textbf{Data-driven approaches} have recently emerged as the most popular approach for air quality forecasting. This research line uses parameterized models, such as neural networks, to capture the spatio-temporal dependencies within air quality data. As opposed to physics-based models, they require far less complex domain knowledge and are usually more flexible. Typically, \citet{zheng2015forecasting} developed a hybrid data-driven model that ensembles the prediction results from different views. Leveraging the large capacity of deep neural networks, DeepAir outperformed existing shallow models on both short and long-term predictions \cite{yi2018deep}. Some follow-ups proposed either STGNNs \cite{wang2020pm25} or attention-based models \cite{wang2021modeling,wang2022air} to better address the spatio-temporal dependencies. These works, however, present some difficulties (e.g., inefficiency, degenerated performance) on a nationwide forecasting task. Besides, there are a stream of studies \cite{pan2019urban,li2020autost,liu2021spatio,pan2021autostg,shao2022pre} exploring new learning paradigms for spatio-temporal data.

\vspace{-0.2em}
\section{Conclusion and Future Work}
We have devised a transformer model for nationwide air quality prediction in China. To the best of our knowledge, this is the first work for collectively forecasting air quality among thousands of locations. Our model elaborately combines the spatio-temporal learning capabilities of transformers with the uncertainty measurement of stochastic latent spaces. Compared to prior methods, our model reduces the prediction errors by 4.6-8.2\%. In the future, we will explore online learning and deploy our model to support public use. 

\section{Acknowledgments}
This research was supported by the MSIT (Ministry of Science, ICT), Korea, under the ITRC (Information Technology Research Center) support program (IITP-2022-2020-0-01789) supervised by the IITP (Institute for Information \& Communications Technology Planning \& Evaluation). It was also supported by the National Natural Science Foundation of China (62172034) and the Beijing Nova Program (Z201100006820053).

\bibliography{aaai23}
\clearpage

\end{document}